# Ultra-low doping and local charge variation in graphene measured by Raman: experiment and simulation

*Zhuofa Chen†, Nathan Ullberg‡, Mounika Vutukuru†, David Barton† and Anna K Swan\*†‡+*

†Department of Electrical and Computer Engineering, Boston University, 8 St Mary's St, Boston Massachusetts 02215, United States of America.

‡Department of Physics, Boston University, 590 Commonwealth Ave, Boston Massachusetts 02215, United States of America.

+Photonics Center, Boston University, 8 St Mary's St, Boston Massachusetts 02215, United States of America.

TOC

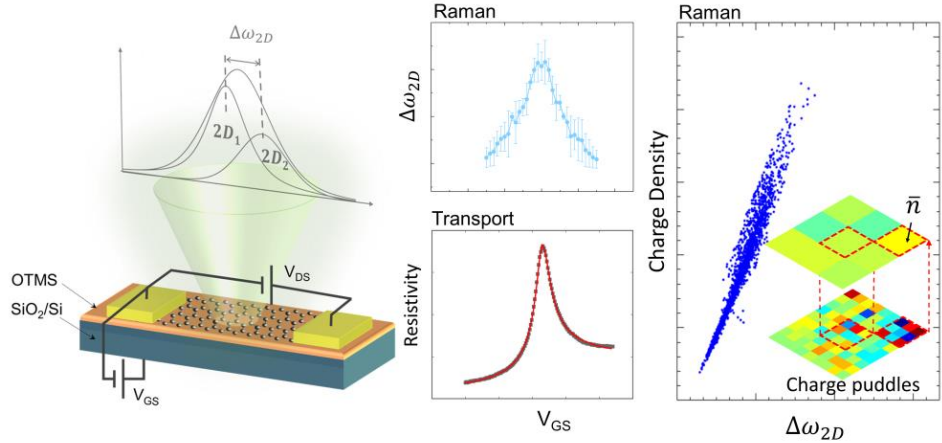

ABSTRACT: Avoiding charge density variations and impurities in graphene is vital for high-quality graphene-based devices. Here, we demonstrate an optical method using Raman 2D peak-split to monitor charge density variations in the range 1-25 × $10^{10}$ cm$^{-2}$. We compare Raman signatures with electrostatically gated Raman and transport measurements to correlate the 2D peak-split with the charge density on graphene with high precision. We found that the Raman 2D peak-split and peak areas linearly varies with the charge density, where a lower charge density results in a larger 2D peak-split. We simulate Raman 2D spectra under various doping conditions to study the correlation between Raman 2D peak and charge puddles. These simulations give qualitative agreement between a sample's measured Raman response and transport properties in graphene. Our work provides a simple and non-invasive optical method for estimating the doping level, local charge density variation and transport properties of graphene before fabricating graphene devices, with up to two orders of magnitude higher precision than previously reported optical methods.

Since its realization in 2004, graphene has attracted a lot of attention due to its superior transport properties such as its giant intrinsic mobility and distinctive electronic structure.[1, 2] The gate-tunable electrostatic charge doping possible for 2D materials, allows for doping without introducing scattering from dopant ions. For the case of graphene, the absence of scattering sites makes for extraordinarily high mobility, which renders graphene an intriguing material for next-generation nanoelectronic devices,[3, 4] including the recently reported unconventional superconductivity in twisted-bilayer graphene that has introduced twistronics as a new platform for applications of graphene.[5] For all high-quality graphene-based devices, reducing charge impurities and charge variations, indicative of charged scattering centers, is essential. Therefore, the identification of charge density variations and impurities in graphene has become vital for studying graphene-based applications.

The benchmark method to determine the quality of a device is the electrical transport measurement. The mobility, the intrinsic charge doping level, and the overall charge density variations can be determined with high precision.[6-8] However, transport measurements require a labor-intensive fabrication process, that may in turn alter the mobility of the material. Our objective is to develop an effective optical method to determine both the doping level and the local charge density variations in graphene before any fabrication process.

Historically, $SiO_2$ has been the most common substrate used for graphene-based devices. Graphene deposited directly on $SiO_2$ has high accidental doping levels ($> 10^{12} cm^{-2}$)[9] and charge puddles formed by charged surface states and impurities in the oxide.[10, 11] Raman spectroscopy has been used as an efficient method to monitor the high (accidental) charge density in graphene on $SiO_2$ substrates above $\sim \pm 10^{10}$ cm$^{-2}$ using the Raman G and 2D peak width, intensity and frequency.[12-18] The above methods are only suitable for evaluating graphene with a doping level higher than $\sim 10^{12}$ cm$^{-2}$.[19-22] For higher purity samples, e.g., suspended graphene,[23] graphene encapsulated between hexagonal boron nitride (hBN) layers,[7] and graphene deposited on octadecyltrimethoxysilane (OTMS)[24], more precise optical methods are required to estimate the doping level of graphene. Here, we demonstrate an optical method that evaluates both the doping level and charge variation in graphene in the range from $10^{10}$ cm$^{-2}$ to $3 \times 10^{11}$ cm$^{-2}$ by using the split in the Raman 2D peak, which appears at low doping levels.[25-28]

The 2D peak in the Raman spectra of graphene arises from the double resonance mechanism of two phonons near the K-points between two nearby Dirac cones.[12, 28, 29] The resonant property makes it sensitive to any perturbation in the electronic states[12, 30] and the phonon dispersion.[15] Since charge screening will affect the electron Fermi velocity, and reduce the Kohn anomaly in the phonon dispersion, the 2D mode can be used to monitor the doping level in graphene. Earlier work has established that the 2D peak becomes asymmetric in the absence of charge, e.g., for suspended graphene and encapsulated graphene.[25, 26, 31, 32] Berciaud et al. investigated the asymmetric 2D line shape of suspended graphene with low charge density ($< 5 \times 10^{11} cm^{-2}$) and found that the 2D peak can be separated into 2 peaks ($2D_1$ and $2D_2$) for low doping.[25,

[26] The origin of 2D$_1$ and 2D$_2$ is ascribed to the inner and outer process of the electron (hole) phonon scattering from two nearby Dirac cones due to asymmetry in the electron and phonon dispersions.[33-36] For graphene sandwiched between hBN layers, the 2D peak-split at low accidental doping reveals a host of information on charge density screening of the electron and phonon dispersions in graphene.[37]

In this work, we use Raman measurements and simulated Raman responses based on gated Raman measurements, combined with electronic transport measurements on the same OTMS-treated SiO$_2$/Si substrates. We compare the Raman response with the transport data which reveals the average global charge density and charge density variations. The electrostatically gated Raman is used to correlate the 2D peak asymmetry with a known charge density. The properties of interest originate from the 2D peak asymmetry, here quantified by two peaks (2D$_1$ and 2D$_2$), their frequency separation (2D peak-split $\Delta\omega_{2D}$) and integrated intensity (peak areas $A_{2D_1}$, $A_{2D_2}$). We find that the Raman 2D peak-split is not continuous, but fall either in a high-split, or low-split regime. Our simulated Raman spectra reveal that the low-split regime is due to the high charge density variation within a laser spot. Our work provides a simple, noninvasive optical method to determine low doping levels and charge variations in graphene with high precision, allowing evaluation of the graphene quality before fabricating graphene-based devices.

**Results and discussion**

**Graphene Raman 2D peak-split**: The Raman spectra for spatial mapping of the graphene samples were measured using a green laser (532 nm) since the 2D peak asymmetry is accentuated for this wavelength in suspended graphene[25] and hBN sandwiched graphene[37] as well as on our graphene on OTMS/SiO$_2$ samples. Spatial Raman maps (Fig. S1) of graphene on OTMS-treated substrates demonstrates slight compression of the graphene sample.[9, 17, 38] Fig. 1a shows the correlation between $\omega_G$ and $\omega_{2D}$, i.e., the "strain line"[17] with a slope of $2.07 \pm 0.08$, where we have used a single peak to fit the 2D phonon (Fig. S2). The G peak is centered around $1585 \pm 0.4$ cm$^{-1}$, a 3 cm$^{-1}$ upshift from the unstrained value, indicating less than 0.1% of compressive strain in this graphene sample.[9, 17] The small compression rules out a 2D peak asymmetry due to strain.[39, 40] The inset in Fig. 1a shows a histogram of the G band linewidth $\Gamma_G = 13.8 \pm 0.24$ cm$^{-1}$. The large value of $\Gamma_G$ indicates that the sample has a doping level below $0.5 \times 10^{12}$ cm$^{-2}$. Clearly, the $\omega_G$ vs. $\omega_{2D}$ charge vector cannot distinguish between charge variations in a low doping regime.[17] Fitting the 2D phonon with 2 peaks reveals that two points that are virtually identical in the $\omega_G$ vs. $\omega_{2D}$ plot in Fig. 1a, show different 2D peak asymmetry (Fig. 1b-c). We will therefore exploit the variations in the 2D peak asymmetry. We fit the 2D peak with two Voigt profiles (see Methods and Fig. S3-5). The lower energy peak we denote as 2D$_1$ (black curve), and the higher energy peak as 2D$_2$ (blue curve), as shown in Fig. 1b-c. For the measurements and analysis to follow, we will use the peak distance between 2D$_1$ and 2D$_2$ to define the 2D peak asymmetry, namely the 2D peak-split, $\Delta\omega_{2D} = \omega_{2D_1} - \omega_{2D_2}$, (a negative value), and the integrated intensity of the two peaks, $A_{2D_1}$ and $A_{2D_2}$. The 2D peak split was found to have a much higher sensitivity to charge density than other optical methods including $\omega_G$, $\Gamma_G$, $\omega_{2D}$, $\Gamma_{2D}$, $I_{2D}/I_G$ (Fig. S6).

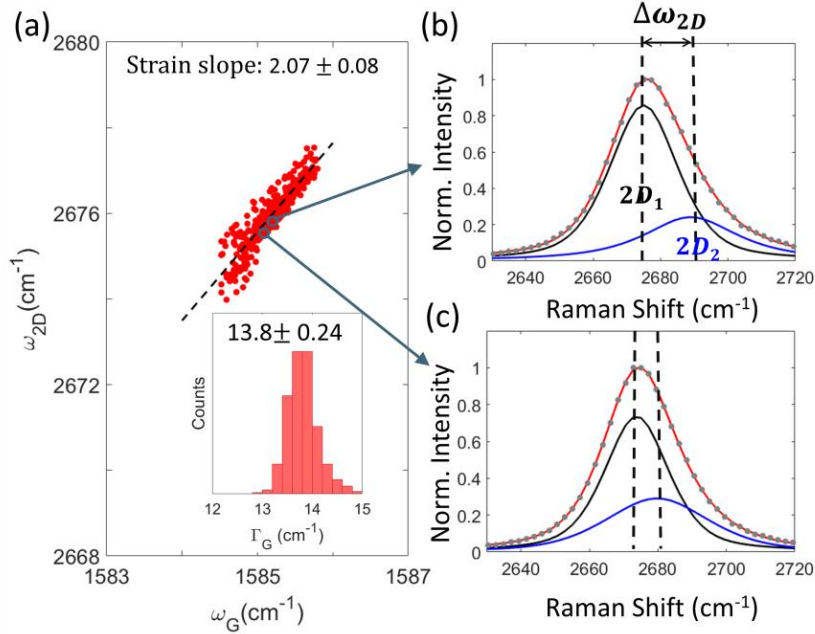

*Fig. 1 The strain-line and Raman 2D peak-split. (a) Strain-line: Correlation between Raman G peak frequency ($\omega_G$) and 2D peak frequency ($\omega_{2D}$) fitted with a single peak demonstrating minimal compression and high homogeneity. The inset shows a histogram of G peak width ($\Gamma_G$). The Raman 2D peak fitting results of two spectra close to each other in (a) are shown in (b) and (c). The 2D peak is fitted by two Voigt profiles, 2D$_1$ (black curve) and 2D$_2$ (blue curve). The 2D peak-split is the difference between 2D$_1$ and 2D$_2$ peak frequency, denoted by $\Delta\omega_{2D} = \omega_{2D_1} - \omega_{2D_2}$.*

**Distribution and properties of 2D peak-split**. Fig. 2a plots the 2D$_1$ peak frequency (black), the 2D$_2$ peak frequency (green and blue), and the 2D peak (single peak fit, red) versus the G peak frequency, i.e., the strain lines. The 2D$_1$ data (black circles) is almost identical to the 2D single fit ($\omega_G$ vs. $\omega_{2D}$ red circles), albeit 1.5 cm$^{-1}$ lower, and has a steeper strain slope of 2.24. The 2D$_2$ data (blue and green) has more variation and thus is more sensitive to charge variation (Fig. S7). Two distinct distributions in the 2D$_2$ vs. G peak frequency emerges: higher and lower energy bands, 2D$_{2+}$ (blue) and 2D$_{2-}$ (green). We have found similar distributions in the 2D$_2$ peaks of BN/Graphene/BN stack samples.[37] We will show below in our simulations that the lower peak splits (2D$_{2-}$) are due to a larger number of charge puddles in the laser spot, even though they may have similar average charge as the 2D$_{2+}$ data points. Fig. 2b shows the spatial map of $\Delta\omega_{2D}$, where the color distribution shows the spatial pattern of the two distinct distributions in $\Delta\omega_{2D}$. Fig. 2c shows the histogram of the $\Delta\omega_{2D}$, 2D$_{2-}$, centered around $\Delta\omega_{2D}$ = -5 cm$^{-1}$ (green), and 2D$_{2+}$, centered around $\Delta\omega_{2D}$ = -14 cm$^{-1}$ (blue). The latter corresponds to $n = \sim 10^{11}$ cm$^{-2}$ as seen by our field gated Raman measurements, shown in the next section. High split data (blue), correlates with low charge density and low charge variation, while data points that fall in the low-split data is a sign of charge variation within the laser spot.

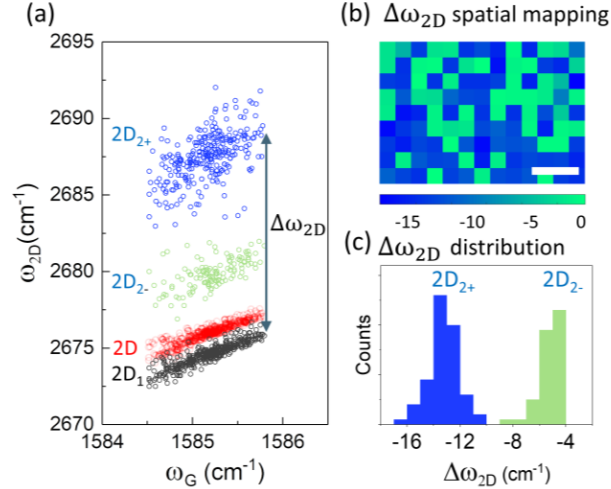

*Fig. 2 Distribution of 2D peak-splits. (a) G peak frequency versus the 2D$_1$ peak frequency, the 2D$_2$ peak frequency, and the single peak fit 2D peak frequency. (b) A Raman spatial mapping of $\Delta\omega_{2D}$. The scale bar in (b) is 3 μm. (c) A histogram of the $\Delta\omega_{2D}$ distribution.*

**Charge density and 2D peak-split.** Next, we directly correlate the Raman $\Delta\omega_{2D}$ with charge by using gated Raman measurements, shown in Fig. 3. Fig 3a shows the schematic of a GFET structure and the conceptual view of the GFET monitored by the Raman 2D peak-split as a function average charge ($n$) close to the charge neutrality point (CNP) controlled by the back gate voltage, Fig 3b. The CNP is determined by two-terminal transport measurements. A transport curve corresponding to the blue data in Fig 3b is shown in Fig 3c, with the CNP located at 1.55V which corresponds to $\bar{n} = 3.6 \times 10^{11}$ cm$^{-2}$. The asymmetric transport curve is typical when a PN junction is formed around the contact region.[41, 42] The fitted mobility $\mu$ is $\sim 17 \times 10^3$ cm$^2$/V·s and the fitted charge variation $\Delta n$ is $2.2 \times 10^{11}$ cm$^{-2}$ (Fitted methods shown in SI and Fig. S8). The doping level and the charge variation are an order of magnitude lower than graphene samples deposited directly on SiO$_2$.[43, 44]

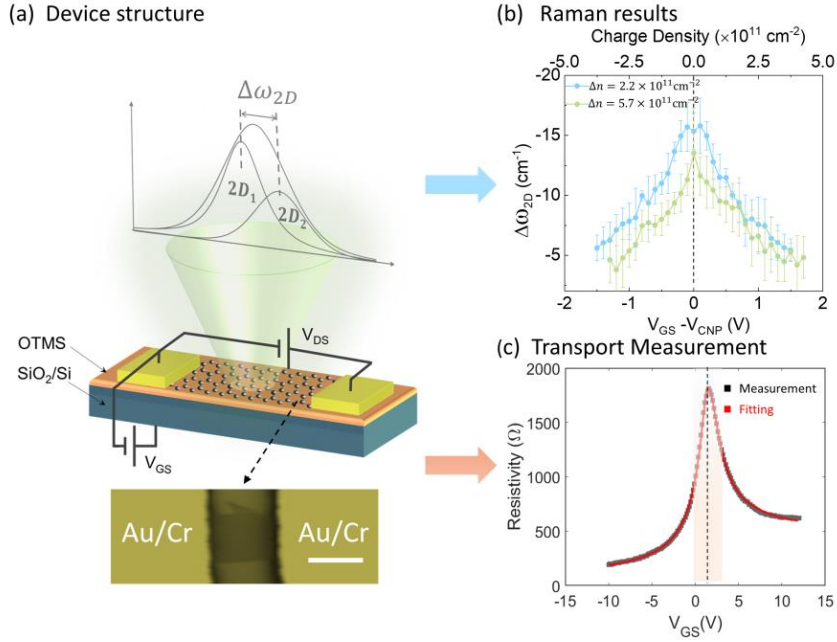

*Fig. 3 Monitoring graphene doping levels by the 2D peak-split. (a) Experimental setup of the GFET monitored by the Raman 2D peak-split. The top inset shows the measured Raman spectrum analyzed by the 2D peak-split. The bottom inset shows an image of a representative GFET device. Scale bar: 10 µm. (b) 2D peak-split vs. back-gate voltage clearly demonstrating how the 2D peak-split increases closer to the CNP. The top axis shows the corresponding charge density. Δn for the blue and green curves are 2.2 and $5.7 \times 10^{11}\ cm^{-2}$, respectively. (c) Transport measurement of GFET, µ = 17 ×10³ cm²/Vs, Δn = 2.2 × $10^{11}\ cm^{-2}$. The shaded region is the low doped regime (from 0 to 3V) used for the gated Raman measurement.*

The gated Raman measurements, Fig. 3b, shows the correlation between the $\Delta\omega_{2D}$ versus the back-gate voltage in a narrow doping range, $\pm\ 3.6 \times 10^{11}\ cm^{-2}$ ($\pm$ 1.5 V on the back gate) close to the CNP. The blue and green denotes data from two devices with different charge variation, measured by transport, where the blue data is from the cleaner sample with lower $\Delta n$. Both plots show that $\Delta\omega_{2D}$ increases as the doping level decreases for both hole and electron doping. We use this behavior to evaluate the charge density of graphene using the Raman $\Delta\omega_{2D}$ around the CNP. By linearizing $\Delta\omega_{2D}$ versus $n$ in the low doping regime from the cleaner sample ($|\Delta\omega_{2D}| > 8cm^{-1}$), we find a variation of $2.3 \times 10^{10}\ cm^{-2}$ per 2D peak-split wavenumber. However, the $\Delta\omega_{2D}$ response to doping is not identical for the two samples shown in Fig. 3. The sample with smaller global Δn consistently shows higher split values than the sample with higher global charge density variation. This behavior can be understood by considering that the 2D peak Raman response is the same for both electron and holes, Fig. 3b. Considering a sample at the CNP but with non-zero local charge puddles in the laser spot, $\bar{n}_{p0} = \delta n_p^+ + \delta n_p^- = 0$, the Raman response will correspond to a charge density that is the average of $|\delta n_p^+| + |\delta n_p^-| \neq 0$, rather than $\bar{n}_{p0} = 0$. Hence, local charge variations will always diminish the $\Delta\omega_{2D}$ at the charge neutrality point.

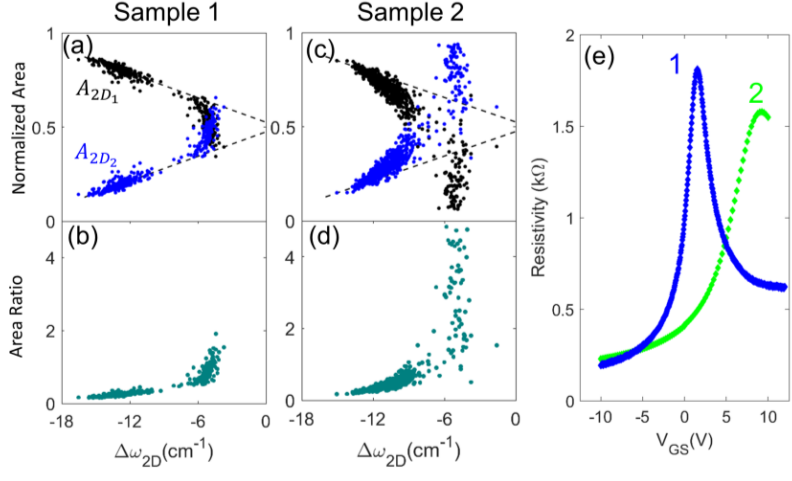

*Fig. 4 Raman and Transport data for two different graphene samples. (a,c) Correlation between the normalized integrated Raman 2D peak intensity area of $A_{2D1}$ (black dots) and $A_{2D2}$ (blue dots) versus $\Delta\omega_{2D}$. The dashed lines show the fit to the high split regime of sample 1 associated with minimal local charge variations within the laser spot. (b,d) The ratio of $2D_2$ peak area over the $2D_1$ peak area is plotted versus $\Delta\omega_{2D}$. (c) The transport results of samples 1 and 2. The fitted transport parameters are $\mu = 17 \times 10^3$ cm$^2$/Vs, $\Delta n = 2.2 \times 10^{11}$ cm$^{-2}$ for sample 1 and $\mu = 5.2 \times 10^3$ cm$^2$/Vs, $\Delta n = 5.7 \times 10^{11}$ cm$^{-2}$ for sample 2.*

The statistics from the Raman maps of the integrated Raman 2D peak intensities $A_{2D_1}$ and $A_{2D_2}$ versus $\Delta\omega_{2D}$, is shown in Fig. 4a,c for two different devices. The corresponding transport measurements are shown in Fig. 4e. The resistivity versus gate voltage show that sample 1 has the highest mobility, ($\mu = 17 \times 10^3$ cm$^2$/Vs) and the lowest global charge variation ($\Delta n = 2.2 \times 10^{11}$ cm$^{-2}$). Sample 2 has lower quality ($\mu = 5.2 \times 10^3$ cm$^2$/Vs, $\Delta n = 5.7 \times 10^{11}$ cm$^{-2}$). The fitting model for transport measurements and the extraction of mobility, accidental doping, and charge variation is described in SI and Fig. S8. The correlation between peak width and 2D peak-split of these samples are shown in Fig. S9. We first consider the cleanest sample, shown in Fig 4a. The data points fall in two groups, high and low peak splits region. The high split region exhibit a linear relation between the area intensities and split, fitted in Fig. 4a top by the black dashed lines for $|\Delta\omega_{2D}| > 10$ cm$^{-1}$, while the data points with $|\Delta\omega_{2D}| < 8$, in the low-split regime do not follow the fit lines. Another way of highlighting the different behavior of the high and low-split region is using the peak area ratio $A_{2D_1}/A_{2D_2}$. Fig 4b shows well-defined lines with different slopes for the high and low peak-split regions. Thirdly, there is a clear gap between the high and low peak-split regions, with very few data points between -10 to -7 cm$^{-1}$. As the sample quality decreases as in sample 2, the Raman data no longer exhibit the linear variation. The crossover point between $A_{2D_1}$ and $A_{2D_2}$ moves towards higher splits, and the area ratio is no longer so well defined in the high-split region and exhibits a much higher area ratio in the low-split region, Fig. 4d. In order to understand the different behavior of the low and high split data points, we carried out simulations to study the effect of local charge puddles on the 2D peak-split and the correlation between 2D peak-split and charge density within the laser spot.

**Simulation of charge puddles and 2D peak-split**. Scanning tunneling microscopy measurements have shown that the charge puddle size can vary significantly, and that it is dependent on the density of impurity ions in the oxide, as well on the overall charge density, with puddle dimensions ∼ 6-100 nm,[43-45] much smaller than the spot size of the laser (~ 400 nm). For each of our measurement, charge puddles of different sizes are sampled within the laser spot, producing a composite Raman spectrum. Hence, we need to consider local charge variation within the laser spot ($\bar{n}_{p0} \pm \delta n_p$), as well as global variations between laser spots ($\bar{n} \pm \Delta n$), where $n$ and $\Delta n$ refers to the global average charge and charge variation of the whole sample. In Fig. 5 we illustrate the linear relationship between charge and the Raman response of the 2D areas versus $\Delta\omega_{2D}$ for a uniform charge, and how it changes when several charge puddles are included in the laser spot. The charge distribution is generated by a Gaussian distribution, $n = \bar{n} \pm \Delta n$ (Fig. S10) and the color of each grid in Fig. 5a represents the charge density ($n_0$) of each charge puddle. The generated Raman spectra results that are shown in Fig. 5b,c assumes one constant charge density (number of charge puddles, $q = 1$) within the laser spot, which reproduces the linear model from the measured charge vs. $\Delta\omega_{2D}$ relation from the gated Raman, Fig. 3b, and the $\Delta\omega_{2D}$ vs. 2D peak area shown in Fig. 4a. The relation between $n$ and the peak areas versus $\Delta\omega_{2D}$ are given by $n = k \cdot (\Delta\omega_{2D} - \Delta\omega_{2Dmax})$, and $A_{2D_1,2D_2} = \frac{1}{2}\left(1 \pm (1 - 2a)\frac{\Delta\omega_{2D}}{\Delta\omega_{2Dmax}}\right) \pm a$. The fitted values are $k = 2.3$ cm$^{-2}$/cm$^{-1}$ with $n$ in units of $10^{10}$ cm$^{-2}$, and $a = 0.03$.

In Fig. 5d-f we consider the case of several charge puddles $q$ in the laser spot, (red dashed square, here $q = 9$ for illustration) which yields an average charge, $\bar{n}_{p0}$, and a resulting charge variation $\delta n_p$ in each laser spot. A varying number of puddles $q$ (here 6 -12) are used to simulate the charge variation in the many laser spots. The resulting Raman spectrum is generated by summing up the $q$ spectra in the laser spot, and fitting the resulting spectrum with two Voigt profiles to extract the area of 2D$_1$ and 2D$_2$ for each laser spot. Fig. 5e shows the distribution of area of 2D$_1$ and 2D$_2$ from the simulated data. Fig. 5f shows the distribution of area ratios $A_{2D_1}/A_{2D_2}$. The simulation results show qualitatively similar distribution as the experimental data, Fig 4a-d, with identifiable high and low-split regime behaviors. The data in the high split regime $|\Delta\omega_{2D}| > 10$ cm$^{-1}$, is a slightly smeared version of the linear model shown in Fig 5b,c.

However, in the low-split regime, the situation is very different, illustrated by the Raman spectra from two laser spots which have the same charge density, but different $\delta n_p$, labeled A and B. We find that while the charge density is low enough to correspond to the high split regime, spectrum B has a larger charge variation, which skews the $2D_2$ peak closer to the $2D_1$ peak, as indicated in Fig 5e,f. The effect of the charge variation reducing the apparent 2D peak-split is understood by considering both the $2D_1$ and $2D_2$ peak frequencies and their intensity behavior as a function of charge. The $2D_1$ intensity dominates near CNP (Fig. 4a,c) and furthermore its spectral position is nearly unaffected by charge (Fig.S7). However, the $2D_2$ peak shifts significantly with charge, shown in Fig. S7, and the peak intensity increases for higher charge. Hence, as a range of charge densities are sampled in a laser spot, the $2D_1$ will have high intensity and be unmoving, while the $2D_2$ will decrease in intensity with increasing $\Delta\omega_{2D}$. The Raman response from a laser spot with two different local charge density variation, shown in Fig 5g,h, demonstrates that the higher charge densities will dominate the $2D_2$ signal and reduce the measured 2D peak-split. Hence, the laser spot A with $q = 9$ and small $\delta n_p$, obeys the linear relationship, but spot B, with larger $\delta n_p$, is found in the low-split regime. Detailed simulation models and methods can be found in SI S6 and Fig. S11.

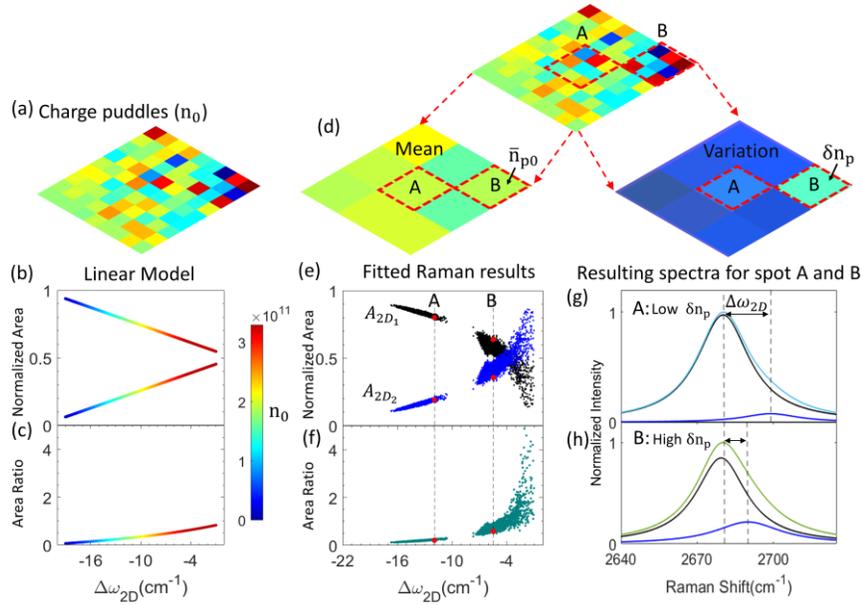

*Fig. 5 Simulated Raman dependence on average charge, and number of charge puddles in a laser spot. (a) Simulated charge puddles, ($\bar{n} = 1 \times 10^{11} \pm 1.2 \times 10^{11} cm^{-2}$). (b-c) Raman results using linear model with one uniform charge per laser spot, q = 1. (d) Several charge puddles per laser spot (red dashed square, q = 9) with average charge density ($\bar{n}_{p0}$) and local charge density variation ($\delta n_p$) calculated by the standard deviation within a laser spot. (e-f) the Raman results using q = 6 - 12, with the same global charge distribution as in (a). The red dots refer to parameters extracted from spectra from spot A and B. (g,h)*

*Calculated 2D peak (light blue), 2D$_1$(black), and 2D$_2$ (blue) from two laser spots (A and B) with the same $\bar{n}_{p0}$ but different $\delta n_p$.*

**Effect of charge puddles**. Fig. 6 compares simulated Raman data with different number of charge puddles $q$ in a lasers spot under three different global charge variations ($\Delta n$). We find that the distribution of $\Delta\omega_{2D}$ not only depends on the local average charge density $\bar{n}_{p0}$ and the charge variation $\delta n_p$, disucssed above, but also the number of charge puddles within a laser spot. For each simulation with a fixed $\Delta n$, we probe the effect on the simulated Raman data as the number of charge puddles increases.

Fig. 6a-i shows the peak areas $A_{2D_1}$ and $A_{2D_2}$ of the simulated Raman data versus $\Delta\omega_{2D}$ with global charge variation $\Delta n$ increasing from left to the right, and number of charge puddles $q$, increasing from top to bottom. The color bar on the top denotes the linear model charge density versus $\Delta\omega_{2D}$. The data-points are colored according to their local average charge density, $\bar{n}_{p0}$. For a nearly homogeneous laser spot ($q$ = 1 - 6, top row), we see that all the data points follow the linear model (dashed line) and moves to lower split $\Delta\omega_{2D}$ and larger spread as the global charge distribution increases, as expected. The most striking observation is that for $q \geq 12$, (Fig 6g-i) all data points end up in the low-split regime, despite that their average charge would put them mostly in the high-split regime. The situation that most closely resembles what we have encountered experimentally is depicted in Fig 6.d-f, where we have data that follows both the linear model in the high-split regime, and data points that fall in the low-split regime despite having $\bar{n}_{p0}$ that belongs in the high split regime.

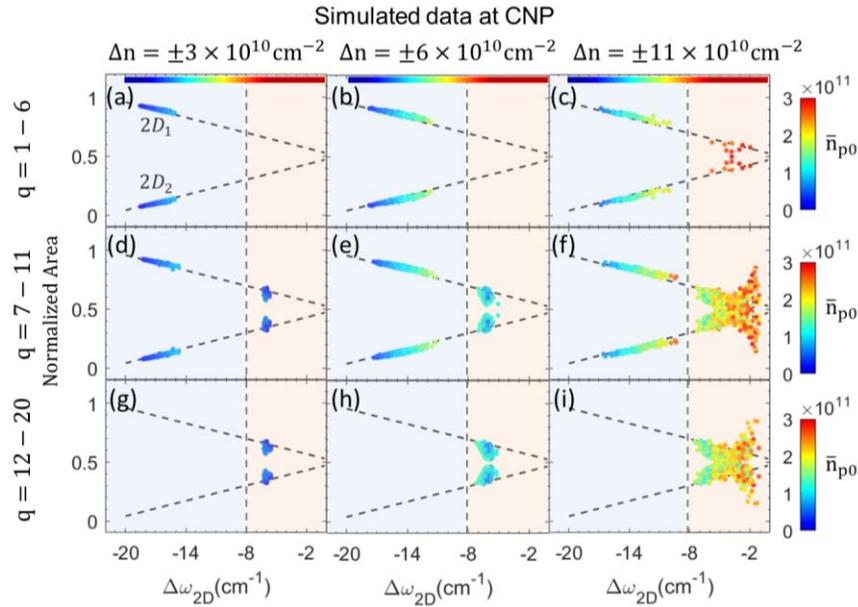

*Fig. 6 Simulated Raman results demonstrating the role of local charge variations, number of puddles, q, in the "laser spot" (top to bottom), and global charge variation n (left to right). The black dashed lines in the*

*figure are the linear fitting model fitted from a clean graphene sample. The top color bars correspond to the charge from the linear model, the color of the data points indicate the $\bar{n}_{p0}$ within a laser spot, as shown on the color bar on the right. The light blue and pink regions indicates the high and low-split regions, respectively.*

The data points (yellow and red) in the low-split region of Fig. 6d-f show that $A_{2D_1}$ reduces and $A_{2D_2}$ increase faster than the linear model as $\bar{n}_{p0}$ increases and $\Delta\omega_{2D}$ reduces. This behavior in the low-split region is emphasized by plotting the area ratio $A_{2D_1}/A_{2D_2}$. Fig. 7a,b shows the area ratio ($A_{2D_1}/A_{2D_2}$) versus $\Delta\omega_{2D}$ ($q$ = 1-20) using the same charge density as in Fig 6 b,c. Fig. 7b resembles our experimental data, where in the high split region (light blue), the area ratio weakly increases as the $|\Delta\omega_{2D}|$ reduces, while in the low-split region (light pink) the slope is very steep and spread out. From the simulated result we can correlate the data points with their average charge, $\bar{n}_{p0}$, the number of puddles, $q$, and their local charge variation $\delta n_{p0}$. Fig. 7c,d shows $\bar{n}_{p0}$ versus $\Delta\omega_{2D}$, which demonstrates that the high split regime (light blue) still shows the linear dependence between charge and split (dashed line corresponds to the linear model used to generate the data). However, in the low-split regime (pink) the split is low, even though the charge density is overlapping with the data in the high split regime. Hence, $\Delta\omega_{2D}$ estimates the local charge density in the high split regime well, but overestimates the charge in the low peak-split regime. However, a larger area ratio indicates a larger average charge density, a trend that can be seen from Fig. 7c,d, where the max area ratio more than doubles as the global charge density doubles.

To summarize the insights from the simulation, we have found that in the high split $\Delta\omega_{2D}$ regime, the average charge density adheres to within 10 % to the linear model. From evaluation of the Fig 7c,d we have as before $n \pm \sigma_n = \text{k} \cdot (\Delta\omega_{2D} - \Delta\omega_{2Dmax})$, where, $\sigma_n$ is the charge density discrepancy, here less than 6 % of $n$. This is further discussed in SI, Fig. S13. Furthermore, the number of charge puddles is relatively low in the high split regime ($q$ is below 12 in our simulation). This shows that the OTMS screens the $SiO_2$ charges sufficiently to increase the linear puddle size to >100 nm. The improvement in screening translates to larger, more homogeneous charge distributions. The data-points in the low-split regime on the other hand, are due to a combination of many more, smaller charge puddles (larger $q$), higher average charge, and higher local charge variation than in the high split regime. It is likely that these areas see increased electron scattering, which negatively affects the mobility due to shorter homogeneous length scales. This assumption is supported by the comparison of the experimental Raman and transport data in Fig. 4. Hence, we can use the mean value and standard deviation of 2D peak-splits to estimate the charge density and charge variation locally in graphene, and hence predict the transport response. High quality graphene will exhibit Raman 2D data with high $\Delta\omega_{2D}$ split in the linear regime, and few or no points in the low-split regime.

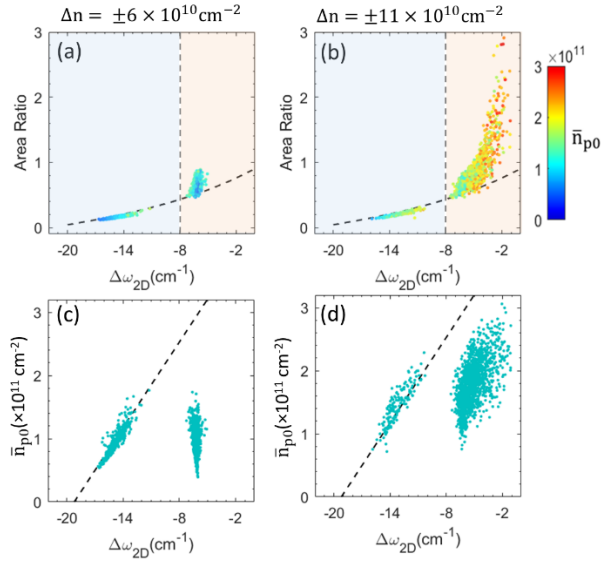

*Fig. 7 Simulated Raman results and transport curves. (a-b) Simulated Raman 2D Area ratio ($A_{2D2}/A_{2D1}$) versus $\Delta\omega_{2D}$ with global charge variation $\Delta n = \pm 6 \times 10^{10} cm^{-2}$ and $\Delta n = \pm 11 \times 10^{10} cm^{-2}$, respectively. The color of the data points represent $\bar{n}_{p0}$, shown on the color bar on the right. (c-d) Average charge $\bar{n}_{p0}$ versus $\Delta\omega_{2D}$ illustrating that the linearity remains in the high-split regime (blue).*

The linearization of the charge to $\Delta\omega_{2D}$ split response is an approximation within limit of validity. Here we have worked in the regime of $1 - 25 \times 10^{10}$ cm$^{-2}$. Furthermore, the constants in the linear model will depend on the dielectric environment as well as the laser frequency used, since the 2D peak not only depends on charge screening, but also on the dielectric screening which affect both the electronic and phonon dispersion.[32] Hence, different a dielectric environment will change the details of the relationship between charge doping and the 2D peak-split. However, the main behavior of the 2D areas and 2D peak split is qualitatively similar, as we have seen in graphene in hBN.[37] Therefore, the 2D peak-split is a sensitive probe of the low doping level, although the absolute values need to be recalibrated for graphene in different dielectric environments. As long as the correlation between low doping charge density and 2D peak-split is recalibrated, one can estimate the low charge density using Raman with high precision without any device fabrication process.

In conclusion, we have demonstrated an optical method to evaluate the low charge density and the low charge density variations across a graphene sample. Using a back-gated GFET, we correlated the 2D peak-split and doping levels and found that the 2D peak-split can differentiate charge densities down to $2.3 \times 10^{10}$ cm$^{-2}$ per 2D peak-split wavenumber, nearly two orders of magnitude higher precision than using G peak frequency and width[18], or the 2D versus G positions[17]. Raman simulations of charge puddles reveals that the two distinct data sets (low and high 2D peak-splits) can be attributed to the number of charge puddles in the laser spot. In the high split regime, the charge density is linearly related to the 2D peak split $\Delta\omega_{2D}$ and has a low number of charge puddles within the laser spot. The low-split regime not only indicate higher charge, but also a higher number of charge puddles. The results show that the statistics of the $\Delta\omega_{2D}$ data (mean and standard deviation) can be used to estimate the average charge and local charge variation in graphene. This method provides a simple, noninvasive way to estimate the doping levels and the quality of a graphene sample before building a high-quality graphene device.

**Methods**

**Sample preparation and device fabrication.** Substrates were first patterned and deposited with back gate contacts. The chips were treated by oxygen plasma to enhance hydrophilicity before the self-assembled monolayer (SAM) was deposited on the $SiO_2$ substrate. Using the spinning and vaporization method for the SAM, 10 μl of OTMS solution (obtained from Sigma– Aldrich) was pipetted on the chip and allowed to settle for 20 s. The chips with OTMS solution were spun at 3000 rpm for 10 s, and subsequently put into a desiccator with ammonium hydroxide solution around the chips for 10 hours which facilitates the formation of the SAM. Finally, the substrates were sonicated in toluene for 5 min to remove residues. A schematic of the device fabrication process can be found in Fig. S14. Atomic force microscopy (AFM) was used to characterize the surface topology, Fig. S15a-b. The smooth surface of the OTMS-treated substrates is achieved by forming a layer of highly ordered OTMS molecules, which reduces dangling bonds and surface-adsorbed polar molecules.[24] The formation of the highly ordered OTMS surface can also be confirmed by contact angle measurements, as in Fig. S15c-d. The highly hydrophobic property indicates the formation of compact hydrophobic groups of the OTMS molecules on the substrate. A comparison of graphene on OTMS surface and SiO2 is shown in Fig. S16.

We use a lithography-free process to fabricate Graphene field - effect transistors (GFETs) on the OTMS-treated substrates. Graphene samples were deposited on the OTMS-treated $SiO_2$/Si chips under ambient conditions using mechanical exfoliation and then identified by optical contrast and confirmed by Raman spectroscopy. Instead of using traditional photolithography which may introduce contamination on the graphene flakes, the source and drain Au/Cr electrodes were deposited by e-beam evaporation using a shadow mask. This was accomplished without touching the graphene sample, details can be found in SI.

**Measurements and data analysis**.

The Raman spectra were measured using a Renishaw Raman instrument with a green laser of 532 nm (2.33 eV). The beam size is 0.4 $\mu$m in diameter. The laser power is set to 2 mW to prevent laser-induced thermal effects on samples during measurement. The exposure time is adjusted to collect > 6k counts for the 2D peak area. A 1200 grooves/mm grating was adopted for the spatial mapping of the Raman response. The collected optical radiation was dispersed onto the charged-coupled device (CCD) array with a spectral dispersion of 2 cm$^{-1}$ per pixel. The IV characteristics of the GFETs were measured by a custom-built setup with two Keithley2400 power sources which were controlled by a MATLAB program for data acquisition and analysis. All of the transport measurements were carried out in a vacuum environment (~10 mTorr) at room temperature.

The 2D$_1$ and 2D$_2$ peaks were fitted with two Pseudo-Voigt profiles (Fig. S3), which is a combination of a Lorentzian and Gaussian profile. To make sure the accuracy and reliability of our analysis, the fitting range of the 2D peak must be at least 300 cm$^{-1}$, which is shown in Fig. S4-5. We also found that the intensity of the spectra should be higher than 3000 counts for stable and reliable fitting results.

ASSOCIATED CONTENT

**Supporting information**

Additional information on substrate preparation and the device fabrication process and measurements, comparing the sensitivity of different Raman parameters due to charge density, comparison of Raman results between graphene on SiO$_2$/Si and graphene on OTMS-treated SiO$_2$/Si substrates, data analysis for fitting range of 2D peak, residual analysis of fitting 2D peaks with a single Voigt and two Voigt profiles, simulation model of Raman 2D peak-split vs. charge density, and the fitting model for transport measurements of GFETs can be found in the SI.

**Corresponding author**

*Email: swan@bu.edu

**Author Contributions**

A.K.S. directed the experimental measurements, Z.C. designed and performed the experiments, Z.C. and N.U. performed Raman data analysis, M.V. built the shadow mask and helped perform Raman measurements, Z.C. and D.B. developed the OTMS substrates coating procedures, A.K.S and Z.C. wrote the manuscript with input from all coauthors.

ACKNOWLEDGMENT

We would like to thank Yuyu Li for fruitful discussion and Ang Liu for the help in device fabrication process. The authors gratefully acknowledge support from the United States National Science Foundation DMR grant 1411008.